\documentclass[onecolumn, 11pt]{revtex4}
\usepackage{graphicx}
\usepackage{amsmath, amsthm, amssymb}
\newtheorem{theorem}{Theorem}

\newtheorem{lemma}{Lemma}

\newcommand{\wwto}{\stackrel{w}{\to}}

\newcommand{\Ran}{\mathrm{Ran}}

\newcommand{\supp}{\text{supp}  }

\newcounter{foo}

\begin{document}

\title{Bound States at Threshold in Systems with Coulomb Repulsion}
\author{Dmitry K. Gridnev}
\email[Electronic address:] {gridnev|at|fias.uni-frankfurt.de}
\affiliation{FIAS, Ruth-Moufang Strasse 1, D--60438 Frankfurt am Main, Germany}
\altaffiliation[On leave from:  ]{ Institute of Physics, St. Petersburg State University, Ulyanovskaya 1, 198504 Russia}
\begin{abstract}
The eigenvalue absorption for a many-particle Hamiltonian depending on a parameter is analyzed in the framework of
non--relativistic quantum mechanics. The long--range part of pair potentials is assumed to be pure Coulomb and no restriction on the particle statistics is imposed. It is proved that if the lowest
dissociation threshold corresponds to the decay into two likewise non--zero charged clusters then the bound state,
which approaches the threshold, does not spread and eventually
becomes the bound state at threshold. The obtained results have a direct application in atomic and nuclear physics. Under minor assumptions a positive proof is given to the conjecture that negative atomic ions have a bound state at the threshold when the nuclear charge becomes critical.
\end{abstract}

\maketitle


\section{Introduction}

In \cite{cmp} using the bounds on two particle Green's function from \cite{we} we analyzed the conditions on pair potentials, which lead to the eigenvalue absorption in the many-body case. For a short list of references concerning the phenomenon of eigenvalue absorption see \cite{exist} (see also \cite{cmp} and references therein). A serious shortcoming of the approach developed in \cite{cmp} was inability to include into consideration the antisymmetry of wave functions, which naturally arises  in the case of fermions. Our aim here is to overcome this difficulty. We shall prove the results similar to those in \cite{cmp} for the pair potentials, which are a sum of a short range part (falling off faster than $r^{-3/2}$) and a pure Coulomb part, and whereby no restrictions on the particle statistics are imposed. This general type of pair potentials covers most important physical cases like atoms, molecules and nuclei. By the end we shall prove rigorously that negative atomic ions have a bound state at threshold when the nuclear charge $Z$ reaches the critical value $Z_{cr}$ and $Z_{cr} > N-1$, where $N$ is the number of electrons.

Similar to \cite{cmp} we consider the Hamiltonian of $N$ particles $H(Z)$, which depends on a parameter $Z \in \mathbb{R}^p$ and has the form
\begin{gather}
    H(Z) = H_0 + V(Z;x) \label{xc31}\\
    V(Z;x) = \sum_{1 \leq i<j \leq N} \left[ U_{ij} (Z; x_i - x_j ) + \frac{q_i (Z) q_j (Z)}{|x_i -x_j|}\right] \label{:xc31}
\end{gather}
where $H_0$ is the kinetic energy operator with the center of mass
removed, $x\in \mathbb{R}^{3N-3}$ is the set of relative coordinates, $x_i \in \mathbb{R}^3$ are particles' position vectors and $q_i (Z) \in \mathbb{R}$ denote the particles' charges, which depend on $Z$. As in \cite{cmp} we consider only $Z \in \mathcal{Z}\subset \mathbb{R}^p$, where the set $\mathcal{Z}$ consists of a given sequence of parameter values $\{Z_k\}_{k=1}^\infty$ converging to some critical
value $Z_k \to Z_{cr}$ and the limit point itself, that is $\mathcal{Z} := \{Z_k\} \cup Z_{cr}$. The pair $(H(Z), \mathcal{Z})$ is called the Hamiltonian with a parameter sequence. The particles are allowed to be bosons and fermions. We denote $\mathcal{P}$ the orthogonal projection operator on the proper symmetry subspace. The bottom of the continuous spectrum we define as $E_{thr}(Z) := \inf \sigma_{ess} (H(Z) \mathcal{P})$.

Throughout the paper we shall use the following function $\eta_\alpha \colon \mathbb{R}^n \to \mathbb{R}$, which determines the asymptotic behavior at infinity $\eta_\alpha (r) = \chi_{\{r|\; |r| \leq 1\}} + \chi_{\{r|\; |r| > 1\}} |r|^{\alpha}$, where $r \in \mathbb{R}^n$ and $\chi_A$ always denotes the characteristic function of the set $A$.
Note that $\eta_\alpha (r)$ is continuous and $\eta_{\alpha_1} \eta_{\alpha_2} = \eta_{\alpha_1 \alpha_2}$. Similar to \cite{cmp} we impose a number of restrictions on the system.
\begin{list}{R\arabic{foo}}
{\usecounter{foo}
    \setlength{\rightmargin}{\leftmargin}}
\item $|U_{ij} (Z; y )| \leq \tilde U (y)$ for all $Z \in \mathcal{Z}, y\in \mathbb{R}^3$, where $\tilde U(y)$ is such that $\eta_\lambda (y) \tilde U(y) \in L^2(\mathbb{R}^3) + L^\infty_\infty(\mathbb{R}^3)$ and $\lambda \in (\frac 32 , 2)$ is a fixed constant. Additionally, there is a constant $q_0 >0$ such that $|q_i(Z) q_j(Z)| \leq q_0$.

\item $\forall f(x) \in C^\infty_0 (\mathbb{R}^{3N-3} )\colon  \lim_{Z_k \to Z_{cr}} \bigl\| \bigl[ V(Z_k) - V(Z_{cr}) \bigr] f \bigr\| =
0$, where $\{Z_k\} = \mathcal{Z}/Z_{cr}$.

\item
for all $ Z_k \in \mathcal{Z}/Z_{cr}$ there are $E(Z_k) \in \mathbb{R}, \psi(Z_k) \in D(H_0)$ such that $H(Z_k) \psi(Z_k) = E(Z_k) \psi(Z_k)$, where $\mathcal{P}\psi(Z_k) = \psi(Z_k)$, $\| \psi(Z_k) \| = 1$ and $E(Z_k) < E_{thr}(Z_k)$.

\item
$\lim_{Z_k \to Z_{cr}} E(Z_k) = \lim_{Z_k \to Z_{cr}} E_{thr}(Z_k) = E_{thr} (Z_{cr})$, where $\{Z_k\} = \mathcal{Z}/Z_{cr}$.
\end{list}

Let $a = 1, 2,
\ldots, (2^{N-1}-1)$ label all the distinct ways \cite{ims} of partitioning
particles into two non--empty clusters $\mathfrak{C}^{a}_1$ and $\mathfrak{C}^{a}_2$. We define the coordinates within the clusters as $\xi^a_{1i}$ and $\xi^a_{2j}$, where $i = 1,2, \ldots, \#\mathfrak{C}^{a}_1$ and $j = 1,2, \ldots, \#\mathfrak{C}^{a}_2$ and the symbol $\#$ refers to the number of particles in the corresponding cluster. The coordinate $\xi^a_{1i} \in \mathbb{R}^3$ points from the center of mass of $\mathfrak{C}^{a}_1$ to the particle $i$ in $\mathfrak{C}^{a}_1$, and $\xi^a_{2j} \in \mathbb{R}^3$ points from the center of mass of $\mathfrak{C}^{a}_2$ to the particle $j$ in $\mathfrak{C}^{a}_2$. By $\xi_a$ we denote the full set of intercluster coordinates consisting of $\xi^a_{1i}$ and $\xi^a_{2j}$. We also notate
\begin{equation}\label{xiamod}
    |\xi_a| = \sum_{i=1}^{\# \mathfrak{C}^{a}_1} |\xi^a_{1i}| + \sum_{j=1}^{\# \mathfrak{C}^{a}_2} |\xi^a_{2j}|
\end{equation}
The coordinate of clusters' relative motion $r_a$ points from the center of mass of $\mathfrak{C}^{a}_1$ to the center of mass of $\mathfrak{C}^{a}_2$.

Following the notation from \cite{reed,ims} we define the sum of interaction cross terms between the clusters as
\begin{equation}\label{14ia}
I_a (Z) := \sum_{\substack{i \in \mathfrak{C}^{a}_1 \\j \in\mathfrak{C}^{a}_2}} V_{ij} (Z)
\end{equation}
The product of net charges of the clusters is defined as
\begin{equation}\label{netcharge}
    Q^a (Z) := \sum_{i \in \mathfrak{C}^{a}_1} \sum_{j \in \mathfrak{C}^{a}_2}  q_i (Z) q_j (Z)
\end{equation}

The projection operators on the proper symmetry subspace for the particles within clusters $\mathfrak{C}_1^{(a)}$ and $\mathfrak{C}_2^{(a)}$ are $\mathcal{P}_1^{(a)}$ and $\mathcal{P}_2^{(a)}$ respectively. Naturally, $\mathcal{P} \mathcal{P}^{(a)}_{1,2} = \mathcal{P}^{(a)}_{1,2} \mathcal{P} = \mathcal{P}$ and $[\mathcal{P}_1^{(a)} ,\mathcal{P}_2^{(a)}] = 0$. We also define $\mathcal{P}^{(a)} := \mathcal{P}_1^{(a)} \mathcal{P}_2^{(a)}$. The Hamiltonian (\ref{xc31}) can be decomposed in the following way
\begin{equation}\label{hthra}
    H(Z) = H^{(a)}_{thr} (Z) - \frac{\hbar^2}{2\mu_a} \Delta_{r_a} + I_a (Z),
\end{equation}
where $H^{(a)}_{thr} (Z)$ is the Hamiltonian of the clusters' intrinsic motion and $\mu_a$ denotes the reduced mass derived from clusters' total masses. From now on without loss of generality we set $\hbar^2/(2\mu_a) = 1$.

It is convenient to use the tensor product space $L^2 (\mathbb{R}^{3N-3}) = L^2 (\mathbb{R}^{3N-6}) \otimes L^2 (\mathbb{R}^{3})$, where the first product term corresponds to the space of $\xi_a$ coordinates and the second term to the space of $r_a$ coordinate. In these terms the operator $H^{(a)}_{thr}$ has the form $H^{(a)}_{thr} = H^{a}_{thr} \otimes 1$, where $H^{a}_{thr}$ is the restriction of $H^{(a)}_{thr}$ to $L^2 (\mathbb{R}^{3N-6})$ (the space of $\xi_a$ coordinates). Similarly, the restriction of $\mathcal{P}^{(a)}$ to the same space we denote by $\mathcal{P}^{a}$, that is $\mathcal{P}^{(a)} = \mathcal{P}^{a} \otimes 1$

The set of requirements continues as follows.
\begin{list}{R\arabic{foo}}
{\usecounter{foo}
    \setlength{\rightmargin}{\leftmargin}}
\setcounter{foo}{4}
\item
For all $Z\in \mathcal{Z}$ and $a=1,2,\ldots,L$ one has $\inf \sigma \bigl(H^a_{thr}(Z) \mathcal{P}^a\bigr) = E_{thr}(Z)$.
\item There is $|\Delta \epsilon| >0$ independent of $Z$ such that the following inequalities hold for all $Z \in \mathcal{Z}$
\begin{gather}
\inf \sigma_{ess} \bigl(H^a_{thr}(Z) \mathcal{P}^a\bigr) \geq E_{thr}(Z) + 2 |\Delta \epsilon| \quad (a=1,\ldots, L) \label{need1}\\
\Bigl[ H^a_{thr} (Z) - E_{thr}(Z)\Bigr] \mathcal{P}^a \geq |\Delta \epsilon| \mathcal{P}^a \quad (a=L+1,\ldots,2^{N-1}-1) \label{need2}
\end{gather}
\end{list}
The requirement R5 says that the bottom of the continuous spectrum of $H(Z)$ is set by the decay into those two clusters that correspond to any of the decompositions $a=1,2, \ldots, L$. Inequality (\ref{need1}) introduces a gap between the ground state energy of the two clusters and other states. For $a=1,2, \ldots, L$ and $Z\in \mathcal{Z}$ we define the projection operator acting on $L^2 (\mathbb{R}^{3N-6}) $ (the space of $\xi_a$ coordinates)
\begin{equation}\label{projopa}
    P^a_{thr} (Z) = \mathbb{P}^a_{(-\infty, E_{thr}(Z) + |\Delta \epsilon|]},
\end{equation}
where $\{\mathbb{P}^a_{\Omega}\}$ are spectral projections of $H^a_{thr}(Z) \mathcal{P}^a$. Note that by R5 one has $\mathbb{P}^a_{(-\infty, E_{thr}(Z))} = 0$ and, hence, by R6 the projection $P^a_{thr} (Z)$ has a finite dimensional range for each $Z$.

We shall need the last requirement, which gives a uniform control over the bound states' wave functions of two clusters.
\begin{list}{R\arabic{foo}}
{\usecounter{foo}
    \setlength{\rightmargin}{\leftmargin}}
\setcounter{foo}{6}
\item
For all $Z \in \mathcal{Z}$ and $a=1,2, \ldots, L$ there are constants $A, \beta >0$ independent of $Z$ and $a$ such that
\begin{equation}\label{perxp}
    \bigl\| e^{\beta |\xi_a|} P^a_{thr}(Z) \bigr\| \leq A,
\end{equation}
where under $e^{\beta |\xi_a|}$ we understand the operator of multiplication by the corresponding function.
\end{list}

The requirement R7 has the following consequence
\begin{lemma}\label{lem:4}
There is an integer constant $\omega >0$ such that
\begin{equation}\label{21}
    \sup_{Z \in \mathcal{Z}} \left[ \dim \Ran P^a_{thr}(Z) \right] \leq \omega
\end{equation}
\end{lemma}
\begin{proof}
$P_{thr}^a(Z) $ is the projection on a finite number of bound states of $H^a_{thr}(Z)$. Thus there must exist orthonormal $\varphi^a_i (Z)\in D(-\Delta) \subset L^2(\mathbb{R}^{3N-6})$ for $i=1,2,\ldots, n(Z)$ such that
\begin{equation}\label{pthranew}
 P_{thr}^a(Z) = \sum_{i=1}^{n(Z)} E_i^a \varphi^a_i (\cdot, \varphi^a_i),
\end{equation}
where the negative numbers $E_i^a (Z)$ lie in the range $[E_{thr}(Z), E_{thr}(Z) + |\Delta\epsilon|]$. We must show that $n(Z) \leq \omega$ for all $Z \in \mathcal{Z}$.
From R7 it follows that
\begin{equation}\label{22}
    (\varphi^a_i , e^{2\beta |\xi_a|} \varphi^a_i ) \leq A \quad  \quad (i = 1,2,\ldots, n(Z))
\end{equation}
for all $Z\in \mathcal{Z}$. From (\ref{22}) we get
\begin{equation}\label{22a}
(\varphi^a_i , \chi_{\{\xi_a |\; |\xi_a| \leq R \}} \varphi^a_i )  \geq \frac 12 ,
\end{equation}
where we define $R:= (\ln{2A})/(2\beta)$. Because the functions $\Delta \varphi^a_i (Z)$, where $\Delta$ is a Laplacian defined on $L^2 (\mathbb{R}^{3N-6})$, are norm bounded uniformly in $Z$ (cf. Lemma~2 in \cite{cmp}) there is a constant $T >0$ independent of $Z$ such that
\begin{equation}\label{23}
-\bigl(\varphi^a_i (Z), \Delta \varphi^a_i (Z)\bigr)  < T
\end{equation}
for all $Z\in \mathcal{Z}$. Combining (\ref{22a}) and (\ref{23}) yields
\begin{equation}\label{24}
\Bigl( \varphi^a_i , \left[ - \Delta - 2T\chi_{\{\xi_a |\; |\xi_a| \leq R \}} \right]\varphi^a_i \Bigr)  < 0
\end{equation}
Because $\varphi^a_i$ are orthonormal by the min--max principle \cite{reed} the value of $n(Z)$ does not exceed the number of negative eigenvalues of the operator in square brackets. This operator does not depend on $Z$ and the number of its bound states having negative energy is finite (which follows, for example, from the Cwikel--Lieb--Rosenblum bound \cite{reed,cwikel}). \end{proof}

So far we have defined $P^a_{thr}(Z)$ on $L^2(\mathbb{R}^{3N-6})$ (the space of $\xi_a$ coordinates). Its extension to the tensor product space we denote as $P^{(a)}_{thr}(Z) = P^a_{thr}(Z) \otimes 1$. We shall need the following simple lemma
\begin{lemma}\label{lem:5}
Suppose that a sequence $f_n \in L^2(\mathbb{R}^{3N-6}) \otimes L^2(\mathbb{R}^{3})$ is uniformly norm--bounded and does not spread. Suppose additionally that an operator sequence $A_n \colon L^2(\mathbb{R}^{3N-6}) \to L^2(\mathbb{R}^{3N-6}) $ is such that $\sup_n \| e^{\alpha |\xi_a|} A_n \| < K$, where $K, \alpha > 0$ are constants. Then the sequence $(A_n \otimes 1 )f_n $ does not spread.
\end{lemma}
\begin{proof}
For the definition of spreading see \cite{cmp}. We can define the full set of relative coordinates for a given cluster partition as $x = (\xi_a, r_a)$ and $|x| = |\xi_a| + |r_a|$. Let us choose $R >0$ so that the following inequalities hold
\begin{gather}
\sup_x \left[ \chi_{\{ x|\; |\xi_a| \geq R\}} e^{-\alpha |\xi_a|} \right]  < \varepsilon/(2K\sup_n\|f_n\|) \label{r1}\\
\left\| \chi_{\{ x|\; |r_a| \geq R \}} f_n \right\| < \varepsilon/2,   \label{r2}
\end{gather}
where $\varepsilon >0$ is some constant. Note that
\begin{equation}\label{chigl}
    \chi_{\{ x|\; |x| \geq 2 R\}} \leq \chi_{\{ x|\; |\xi_a| \geq R\}}  + \chi_{\{ x|\; |r_a| \geq R\}}
\end{equation}
Using (\ref{chigl}) and (\ref{r1})--(\ref{r2}) we obtain
\begin{equation}\label{chigl2}
    \Bigl\| \chi_{\{ x|\; |x| \geq 2 R\}} (A_n \otimes 1) f_n \Bigr\| < \varepsilon
\end{equation}
for all $n$. This proves the claim.  \end{proof}

Now we can formulate the main theorem.
\begin{theorem}\label{th:main}
Suppose that $(H(Z), \mathcal{Z})$ satisfies the requirements $R1-7$ and $Q^{1\leq a \leq L}(Z) > Q_0$ for all $Z\in \mathcal{Z}$, where $Q_0 >0$ does not depend on $Z$. Then (a) for $Z_k \to Z_{cr}$ the sequence $\psi(Z_k)$ defined by R3 does
not spread. (b) $H(Z_{cr})$ has at least one bound state, which is invariant under $\mathcal{P}$ and has the energy $E_{thr}(Z_{cr})$.
\end{theorem}
We postpone the proof to Sec.~\ref{sec:proofitself}. A few remarks are in order. For $L=1$ and particles that are not fermions Theorem~\ref{th:main} can be considered as a partial case of Theorem~4 proved in \cite{cmp}. In contrast, hereby we do not make any restrictions on the particle statistics: the particles can be bosons or fermions. One can improve Theorem~\ref{th:main} by easing some of the restrictions. For example, the exponential fall off in R7 can be replaced by some power. Using the trick from Lemma~9 in \cite{cmp} one can generalize the above theorem to the case of a multi--cluster decay: one must require that the lowest dissociation threshold corresponds to the decay into likewise non--zero charged clusters (the proof would be given elsewhere).

\section{Upper Bound on the Two Particle Green's Function}

Consider the following integral operator on $L^2 (\mathbb{R}^3)$
\begin{equation}\label{def:Gk}
    G^{c}_k (A) = \left[ -\Delta + A \eta_{-1}(x) + k^2\right],
\end{equation}
where $x \in \mathbb{R}^3$ and $A,k >0$. The kernel of this operator we shall denote as $G^{c}_k (A;x,x')$ (the superscript ``c'' in the expression refers to ``Coulomb''). The following Lemma uses the upper bound on a two particle Green's function from \cite{we}.
\begin{lemma}\label{lemma:1}
For a fixed $A > 0$ and all $n>0$ there is a constant $b >0$ such that
\begin{equation}\label{mainnorm}
    \sup_{k>0} \bigl\| G^{c}_k (A) \chi_{\{ x|\: |x| \leq n\}} \bigr\| \leq bn,
\end{equation}
where $\|\cdot\|$ means an operator norm on $L^2 (\mathbb{R}^3)$.
\end{lemma}
\begin{proof}
The operator $G^{c}_k (A) $ is an integral operator with a positive kernel \cite{kernel} and, hence, it suffices to consider (\ref{mainnorm}) for $n>1$. For a shorter notation we denote $\chi_n := \chi_{\{x|\; |x| \leq n\}}$. The following inequality is obvious
\begin{equation}\label{1}
\| G^{c}_{k}(A)\chi_n \| \leq \| \chi_{4n} G^{c}_{k} (A) \chi_n \| + \|
(1-\chi_{4n}) G^{c}_{k}(A)  \chi_{n} \|
\end{equation}
We consider separately two terms on the rhs of (\ref{1})
to obtain $\| \chi_{4n} G^{c}_{k} (A) \chi_n \| = O(n)$ and
$\| (1-\chi_{4n}) G^{c}_{k} (A) \chi_n \| = o(n)$ for $n \to \infty$. From these relations
the statement follows. Because $G^{c}_{k}$ is an integral operator
with a positive kernel the following bound holds for the first term
in (\ref{1}) $\| \chi_{4n} G^{c}_{k} (A)\chi_n \|  \leq \|
\chi_{4n} G^{c}_{k} (A)\chi_{4n} \|  $. The last expression is the norm
of a self--adjoint operator, which can be rewritten as
\begin{gather}
\| \chi_{4n} G^{c}_{k} (A)\chi_{4n} \| = \sup_{\| f \| =
1} \Bigl( \chi_{4n} f,  G^{c}_{k} (A)\chi_{4n} f \Bigr) \leq \label{3}\\
\sup_{\| f \| = 1} \Bigl( \chi_{4n} f, (A\eta_{-1})^{-1}\chi_{4n} f \Bigr)  = 4 A^{-1} n \label{4}
\end{gather}
where we have used that from the operator inequality for positive
self--adjoint operators $B \geq C $ the inequality $B^{-1} \leq
C^{-1} $ follows. Taking $B = -\Delta + A\eta_{-1} + k^2$
and $C = A \eta_{-1}$ we find that (\ref{3})--(\ref{4}) is true. Thus
we obtain $\| \chi_{4n} G^{c}_{k} (A) \chi_n \| = O(n)$ as promised.

Let us now consider the second term on the rhs of (\ref{1}). We
shall need the bound on the Green's function from \cite{we}. Let $\tilde G_k (a; x,x')$ denote the integral kernel of the following operator on $L^2 (\mathbb{R}^3)$
\begin{equation}\label{5}
\tilde G_k (a) = \left[ -\Delta + \left( \frac{a^2}4 |x|^{-1}  +
\frac{a}4 |x|^{-3/2} \right)\chi_{\{x|\: |x| \geq 1\}} + k^2
\right]^{-1}
\end{equation}
Lets us set $a$ equal to the positive root of the equation $a(a+1) = 4A$. Then we get
\begin{equation}\label{.9}
A \eta_{-1} (x) \geq \left( \frac{a^2}4 |x|^{-1}  + \frac a4|x|^{-3/2} \right)\chi_{\{x|\: |x| \geq 1\}},
\end{equation}
which means that $G^{c}_{k} (A;x,x' ) \leq \tilde G_k (a;x,x' )$ pointwise for all $x,x' $, see \cite{we,cmp}. The upper bound on $\tilde G_k (a; x,x' )$ from \cite{we} (Eqs.(42)--(43) and Eqs.~(39)--(40) in \cite{we}) reads
\begin{equation}\label{7}
\tilde G_k (a;x,x') \leq \frac 1{4 \pi |x-x'|} \times
\left\{
\begin{array}{ll}
    1  & \quad \textrm{for $|x-x'| \leq \tilde R_0 $} \\
    \exp\left\{ \tilde a \sqrt{\tilde R_0 } - \tilde a \sqrt{|x-x'|} \right\} & \quad \textrm{for $|x-x'| > \tilde R_0 $}, \\
    \end{array}
    \right.
\end{equation}
where $\tilde R_0 , \tilde a$ have to be chosen to satisfy the following inequalities.
\begin{gather}
\tilde R_0 \geq 1 + |x'| \label{12}\\
\tilde a \leq a \tilde R^{3/2}_{0} (\tilde R_0 + |x'|)^{-3/2} \label{13}
\end{gather}
From the inequality (\ref{7}) we obtain the bound
\begin{equation}\label{8}
\tilde G_{k} (a;x,x') \chi_{\{ |x'| \leq n \}}\leq \frac 1{4
\pi |x-x'|} \times \left\{
\begin{array}{ll}
    1  & \quad \textrm{for $|x-x'| \leq 2n $} \\
    \exp\left\{ \frac a2 \bigl( \sqrt{2n } - \sqrt{|x -x'|} \bigr) \right\} & \quad \textrm{for $|x-x'| > 2n $}, \\
    \end{array}
    \right.
\end{equation}
where we have set $\tilde R_0 = 2n $ and $\tilde a = a/2$. It is straightforward to check that this choice of $\tilde R_0, \tilde a$ indeed satisfies (\ref{12})--(\ref{13}). Taking into account that $G^{c}_{k} (A;x,x' ) \leq \tilde G_k (a;x,x' )$ we finally get from (\ref{8}) the required bound
\begin{equation}\label{15}
G^{c}_{k} (A;x,x') \chi_{\{ x,x'|\: |x| \geq 4n ,  |x'| \leq n \}} \leq \frac{e^{\frac a2 \left( \sqrt{2n} - \sqrt{|x|- n} \right)}}{4\pi (3n)} \chi_{\{ x,x'|\: |x| \geq 4n ,  |x'| \leq n \}}
\end{equation}
Note that the rhs of (\ref{15}) does not depend on $k$. Using the upper bound (\ref{15}) and estimating the operator norm through the Hilbert--Schmidt norm we get
\begin{gather}
    \| (1-\chi_{4n}) G^{c}_{k} (A)\chi_n \|^2 \leq \int_{|x|\geq 4n} dx
    \int_{|x'| \leq n} dx' |G^{c}_{k} (A;x,x' )|^2  \leq \\
    \frac n{27} e^{a\sqrt{2n}} \int_{4n}^{\infty} e^{-a\sqrt{t-n}} t^2 dt \label{18}
\end{gather}
The integral in (\ref{18}) can be calculated explicitly and we obtain $\| (1-\chi_{4n}) G^{c}_{k} (A) \chi_n \| = o(n)$ as claimed.  \end{proof}

We shall need the following corollary of Lemma~\ref{lemma:1}
\begin{lemma}\label{lemma:2}
For the fixed $A>0, \alpha >3/2$ the following inequality holds
\begin{equation}\label{20}
    \sup_{k>0} \bigl\|  G^{c}_{k} (A)\; \eta_{-\alpha}\bigr\| < \infty
\end{equation}
\end{lemma}
\begin{proof}
For an arbitrary $f \in L^{2} (\mathbb{R}^3 )$ we have
\begin{gather}
\bigl\| G^{c}_{k} (A)\eta_{-\alpha} f \bigr\|  = \lim_{N \to \infty} \Bigl\| \sum_{n =
1}^N G^{c}_{k}  (A) \eta_{-\alpha}
(\chi_n - \chi_{n-1} ) f \Bigr\| \label{31}   \\
\leq \lim_{N\to \infty} \sum_{n = 1}^N \Bigl\| G^{c}_{k}  (A) \chi_n \eta_{-\alpha}
(\chi_n - \chi_{n-1})^2 f \Bigr\|
 \label{32}
\end{gather}
where we have used $(\chi_n - \chi_{n-1})^2 = (\chi_n - \chi_{n-1})$
and $\chi_n (\chi_n - \chi_{n-1}) = (\chi_n - \chi_{n-1})$.
For the operator norms we have $\| \eta_{-\alpha} \chi_1 \| = 1$ and $\|
\eta_{-\alpha} (\chi_n - \chi_{n-1}) \| = (n-1)^{-\alpha}$ for
$n\geq 2$. Substituting these into (\ref{32}) and using Lemma~\ref{lemma:2} we rewrite (\ref{32}) as
\begin{gather}\label{33}
\| G^{c}_{k} (A)\eta_{-\alpha} f \| \leq b \lim_{N\to \infty} \left( \bigl\|\chi_1 f \bigr\| + \sum_{n =
2}^N  n(n-1)^{- \alpha} \bigl\| (\chi_n - \chi_{n-1})f \bigr\|\right)
\end{gather}
Now using that $\sum_n \|(\chi_n - \chi_{n-1} )f \|^2 =\|f\|^2$ and applying the
Cauchy-Schwartz inequality we get from Eq.~(\ref{33})
\begin{gather}\label{34}
\| G^{c}_{k} (A)\eta_{-\alpha} \| \leq b \left( 1 + \sum_{n =
2}^\infty  n^2 (n-1)^{- 2\alpha} \right)^{1/2}
\end{gather}
For $\alpha > 3/2$ the series on the rhs of Eq.~(\ref{34}) obviously
converge and we see that indeed $\| G^{c}_{k} (A) \eta_{-\alpha} \|$ is
bounded by a constant independent of $k$. \end{proof}

\section{Proof of the Main Theorem}\label{sec:proofitself}

We shall use the IMS localization formula, see \cite{ims}. The functions $J_a \in C^2
(\mathbb{R}^{3N-3})$ form the partition
of unity $\sum_a J^{2}_a =1$ and are homogeneous of
degree zero in the exterior of the unit sphere, {\em i.e.} $J_a (\lambda x) =
J_a (x)$ for $\lambda \geq 1$, $|x|= 1$ (this makes $|\nabla
J_a |$ fall off at infinity). Additionally, there exists a constant $C > 0$ such that
\begin{equation}\label{ims5}
    \supp J_a \cap \{ x | |x| > 1 \} \subset \{x|\; |x_i - x_j | \geq C |x| \quad
    \textrm{for $i \in \mathfrak{C}_1^a , j \in \mathfrak{C}_2^a$}\}
\end{equation}
The functions of the IMS decomposition are chosen \cite{sigalsays1,sigalsays2} to be invariant under permutations of particle coordinates both in $\mathfrak{C}_1^a$ and in $\mathfrak{C}_2^a$, hence $[J_a, \mathcal{P}^{(a)}] = 0$.

Following \cite{ims,reed} we introduce
\begin{equation}\label{cycha}
    H_a (Z)= H(Z) - I_a (Z)= H^{(a)}_{thr}(Z) - \Delta_{r_a}
\end{equation}

We shall need the following analogue of the IMS localization formula \cite{ims}, which can be verified by the direct substitution
\begin{equation}\label{14}
    H_0 = H_0 \sum_{a,b} J^2_a J^2_b  = \sum_{a,b} J_a J_b H_0 J_b J_a + 2 \sum_a |\nabla J_a|^2
\end{equation}
The second sum on the rhs of (\ref{14}) is relatively $H_0$ compact \cite{ims}. The whole Hamiltonian can be written as
\begin{equation}\label{ims3}
    H(Z) = \sum_a J^2_a H_a (Z) J^2_a + \sum_{a\neq b} J_a J_b H_{ab}(Z) J_b J_a + K(Z)
\end{equation}
where we define
\begin{gather}
    K(Z) := \sum_{a\neq b} J^2_a J_b^2 I_{ab}(Z) + \left[ \sum_a J^{4}_a I_a(Z) + |\nabla J_a|^2 \right] \label{14k}\\
H_{ab}(Z) := H_0 + \sum_{s=1}^2 \sum_{p=1}^2 \: \sum_{i,j \in \mathfrak{C}^{a}_s \cap \mathfrak{C}^{b}_p} V_{ij} (Z) \\
I_{ab} (Z) := H(Z) - H_{ab}(Z)
\end{gather}
The Hamiltonian $H_{ab}$ defined for $a \neq b$ contains interactions within four clusters $\mathfrak{C}^a_s \cap \mathfrak{C}^b_p$, where $s,p = 1,2$ and all cross--terms between these four clusters are contained in $I_{ab}$. (For some partitions it might happen that one of the four clusters is empty). If we define by $\mathcal{P}^{(ab)}_{sp}$ the projection operator on the proper symmetry subspace for particles within the cluster $\mathfrak{C}^a_s \cap \mathfrak{C}^b_p$ then by the HVZ theorem
\begin{equation}\label{infsigm}
    \inf \sigma \bigl( H_{(ab)}(Z) \mathcal{P}^{(ab)} \bigr) \geq E_{thr} (Z),
\end{equation}
where we define
\begin{equation}\label{pabproj}
    \mathcal{P}^{(ab)} := \mathcal{P}^{(ab)}_{11} \mathcal{P}^{(ab)}_{12} \mathcal{P}^{(ab)}_{21} \mathcal{P}^{(ab)}_{22}
\end{equation}
Note that $[J_a J_b , \mathcal{P}^{(ab)}] = 0$.

\begin{lemma}\label{lemma:3}
Suppose that $(H(Z), \mathcal{Z})$ satisfies R1-7. Suppose, additionally, that $\psi(Z_n) \equiv \psi_n
\wwto \phi_0 \in D(H_0)$, where $Z_n \to Z_{cr}$ and $Z_n \in \mathcal{Z}/Z_{cr}$. Then
\begin{gather}
    \lim_{Z_n \to Z_{cr}} \Bigl\| \left[1- P^{(a)}_{thr}(Z_n)\right] J^2_a (\psi_n - \phi_0) \Bigr\| = 0 \quad (a=1, \ldots, L) \label{claim1}\\
\lim_{Z_n \to Z_{cr}} \Bigl\| J^2_a (\psi_n - \phi_0) \Bigr\| = 0  \quad (a=L+1, \ldots, 2^{N-1}-1)\label{claim1:1}
\end{gather}
\end{lemma}
\begin{proof}
Following the arguments from \cite{cmp} (in the proof of Lemma~8 in \cite{cmp} after Eq.~(61)) one can show that
\begin{equation}\label{k112:}
\lim_{Z_n \to Z_{cr}} \bigl( (\psi_n-\phi_0), \left[ H(Z_n) - E_{thr}(Z_n) \right](\psi_n-\phi_0) \bigr) = 0.
\end{equation}
Similar to \cite{cmp} we define $\tilde K$ by (\ref{14k}), where all $V_{ij}$ entering $K$ are replaced by
\begin{equation}\label{tildevij}
    \tilde V_{ij} := \tilde U_{ij} + \frac{q_0}{|x_i - x_j|} ,
\end{equation}
where $\tilde U_{ij} := \tilde U (x_i - x_j)$.
Then $\tilde K$ does not depend on $Z$ and is relatively $H_0$ compact and besides for all $f \in D(H_0)$ one has $|(f, K(Z)f)| \leq (f, \tilde K f)$. Thus by Lemma~3 in \cite{cmp} we conclude
\begin{equation}\label{k112:2}
\lim_{Z_n \to Z_{cr}} \bigl( (\psi_n-\phi_0), K(Z_n) (\psi_n-\phi_0) \bigr) = 0.
\end{equation}
Substituting (\ref{ims3}) into (\ref{k112:}) and using (\ref{k112:2}) yields
\begin{gather}\label{19}
\lim_{Z_n \to Z_{cr}} \sum_a \Bigl( (\psi_n - \phi_0), J^2_a \left[H_a (Z_n) - E_{thr} (Z_n)\right]J^2_a (\psi_n - \phi_0)\Bigr) + \\
\sum_{a\neq b} \Bigl( (\psi_n - \phi_0) , J_a J_b \left[H_{ab} (Z_n) - E_{thr} (Z_n)\right] J_b J_a (\psi_n - \phi_0)\Bigr) = 0
\end{gather}
Note that all scalar product terms are non--negative. For example, the terms in the second sum are non--negative by (\ref{infsigm}) (one can insert $\mathcal{P}^{(ab)}$ because $\mathcal{P}^{(ab)} \mathcal{P} = \mathcal{P}$ and $[J_a J_b , \mathcal{P}^{(ab)}] = 0$).
Thus we obtain for all partitions $a=1,\ldots, 2^{N-1}-1$
\begin{equation}\label{allparts}
 \lim_{Z_n \to Z_{cr}}    \Bigl( (\psi_n - \phi_0), J^2_a \left[H_a (Z_n) - E_{thr} (Z_n)\right] J^2_a (\psi_n - \phi_0)\Bigr) = 0
\end{equation}
Using (\ref{cycha}) and $-\Delta_{r_a}$ being non--negative gives
\begin{equation}\label{allparts:1}
 \lim_{Z_n \to Z_{cr}}    \Bigl( (\psi_n - \phi_0), J^2_a \left[H^{(a)}_{thr} (Z_n) - E_{thr} (Z_n)\right] \mathcal{P}^{(a)} J^2_a (\psi_n - \phi_0)\Bigr) = 0,
\end{equation}
where we have inserted $\mathcal{P}^{(a)}$. For $a \geq L+1$  the statement of the lemma given by (\ref{claim1:1}) easily follows from (\ref{allparts:1}) and from R6. To prove (\ref{claim1}) it suffices to insert into (\ref{allparts:1}) the identity $1 = P^{(a)}_{thr} + (1-P^{(a)}_{thr})$ and to use the inequality
\begin{equation}\label{insineq}
    \left[H^{(a)}_{thr} (Z_n) - E_{thr} (Z_n)\right] \left(1-P^{(a)}_{thr}(Z_n) \right)\mathcal{P}^{(a)}  \geq |\Delta  \epsilon|\left(1-P^{(a)}_{thr}(Z_n) \right)\mathcal{P}^{(a)} ,
\end{equation}
which follows from (\ref{projopa}). This proves the claim.  \end{proof}

The following lemma expresses the idea of the multipole expansion
\begin{lemma}\label{lem:thet}
There is $\Theta_a (x) \in L^2 (\mathbb{R}^{3N-3}) + L^\infty_\infty(\mathbb{R}^{3N-3}) $ independent of $Z$ such that for all $Z \in \mathcal{Z}$
\begin{equation}\label{thetaa}
    \Bigl| e^{-\beta  |\xi_a|} \eta_\lambda (r_a) \bigl[ I_a (Z) - Q^a (Z)\eta_{-1} (r_a) \bigr] \Bigr|\leq \Theta_a (x)
\end{equation}
\end{lemma}

\begin{proof}
The statement of the lemma is based on the following inequality, which can be checked directly. For all $s, s' \in \mathbb{R}^3$
\begin{equation}\label{paxi}
    \left| \frac{\chi_{\{s,s' |\; |s-s'| \geq 1\}}}{|s-s'|} - \eta_{-1}(s)\right| \leq 2 \eta_{2}(s')\eta_{-2}(s)
\end{equation}
For fixed $s'$ the term on the lhs of (\ref{paxi}) falls off like $|s|^{-2}$. We write
\begin{gather}
\Bigl| I_a (Z) - Q^a (Z)\eta_{-1} (r_a) \Bigr|  \leq \sum_{\substack{i \in \mathfrak{C}^{a}_1 \\j \in\mathfrak{C}^{a}_2}} \tilde U_{ij}  + \sum_{\substack{i \in \mathfrak{C}^{a}_1 \\j \in\mathfrak{C}^{a}_2}} \frac{q_0}{|x_i - x_j|} \chi_{\{x |\; |x_i-x_j| \leq 1\}} \\
+ \sum_{\substack{i \in \mathfrak{C}^{a}_1 \\j \in\mathfrak{C}^{a}_2}} \bigl| q_i (Z) q_j (Z) \bigr| \left| \frac{\chi_{\{x |\; |x_i-x_j| \geq 1\}}}{|x_i-x_j|} - \eta_{-1} (r_a) \right|
\end{gather}
By (\ref{paxi}) we have
\begin{equation}\label{pshe}
    \left| \frac{\chi_{\{x |\; |x_i-x_j| \geq 1\}}}{|x_i-x_j|} - \eta_{-1} (r_a) \right| \leq \eta_2 (|\xi_a|) \eta_{-2}(r_a) ,
\end{equation}
where we have used that $x_j-x_i = \xi^a_j + r_a - \xi^a_i$. Using (\ref{pshe}) we conclude that the inequality (\ref{thetaa}) would be true if we set $\Theta_a = \Theta_{a1} + \Theta_{a2}$, where
\begin{gather}
\Theta_{a1}(x) := e^{-\beta  |\xi_a|} \eta_\lambda (r_a) \sum_{\substack{i \in \mathfrak{C}^{a}_1 \\j \in\mathfrak{C}^{a}_2}} \Bigl[ \tilde U_{ij} + \frac{q_0}{|x_i - x_j|} \chi_{\{x |\; |x_i-x_j| \leq 1\}} \Bigr] \\
\Theta_{a2}(x) := N (N-1) q_0 e^{-\beta  |\xi_a|} \eta_2 (|\xi_a|) \eta_{\lambda-2}(r_a)
\end{gather}
Using R1 it is easy to see that $\Theta_{a1} \in L^2 (\mathbb{R}^{3N-3}) + L^\infty_\infty(\mathbb{R}^{3N-3})$. Because $\lambda < 2$ we have $\Theta_{a2}  \in L^\infty_\infty(\mathbb{R}^{3N-3})$.
\end{proof}

The previous lemma helps proving the following statement, which is the main ingredient in the proof of Theorem~\ref{th:main}.

\begin{lemma}\label{lem:6}
Suppose that the conditions of Theorem~\ref{th:main} are fulfilled. Suppose, additionally, that $\psi(Z_n) \equiv \psi_n \wwto \phi_0 \in D(H_0)$, where $Z_n \to Z_{cr}$ and $Z_n \in \mathcal{Z}/Z_{cr}$. Then for $a=1,2, \ldots, L$ the sequence $P^{(a)}_{thr} (Z_n) \psi_n $ does not spread.
\end{lemma}
\begin{proof}
For a shorter notation let us set $I_a^n := I_a (Z_n)$ and $Q_n^a := Q^a (Z_n)$. The Schr\"odinger equation reads
\begin{gather}
    \Bigl\{ \bigl[ H^{(a)}_{thr}(Z_n) - E_{thr}(Z_n)\bigr ] -\Delta_{r_a}  + Q_n^a \eta_{-1} (r_a ) +
    k_n^2 \Bigr \} \psi_n \label{schro:1}\\
    = - \left[ I^n_a - Q_n^a \eta_{-1} (r_a ) \right] \psi_n , \label{schro:2}
\end{gather}
where we define $k_n^2 = E_{thr}(Z_n) - E(Z_n)$ and also add and subtract the term $Q_n^a \eta_{-1}$. By Lemma~\ref{lem:4} we can write
\begin{equation}\label{parepr}
    P^{(a)}_{thr} (Z) = \sum_{i=1}^\omega E_i^a(Z) P_{\varphi^a_i} (Z),
\end{equation}
where $P_{\varphi^a_i} (Z) = \varphi^a_i (\cdot, \varphi^a_i) \otimes 1$ and some $\varphi^a_i (Z)$ can be zero. Recall that $\varphi^a_i (Z)$ are orthonormal eigenstates of $H^{(a)}_{thr}(Z)$ with the energies $E_i^a (Z)$. For these energy values we have $E_i^a (Z_n) \in [E_{thr}(Z_n) , |\Delta \epsilon|] $. Because the sum in (\ref{parepr}) runs over a finite number of terms to prove the theorem it suffices to show that $P_{\varphi^a_i} (Z_n) \psi_n $ does not spread.

Acting
by $P_{\varphi^a_i} (Z_n)$ on both sides of (\ref{schro:1})--(\ref{schro:2}) we obtain
\begin{gather}
\left[ -\Delta_{r_a} + Q_n^a \eta_{-1} (r_a) + {k'}_n^2 \right]
P_{\varphi^a_i} (Z_n) \psi_n \label{schro:3}\\
= - P_{\varphi^a_i} (Z_n)\left[ I^n_a - Q_n^a \eta_{-1} (r) \right] \psi_n \label{schro:4},
\end{gather}
where ${k'}_n:= \bigl[| E_i^a (Z_n) - E_{thr}(Z_n)| + k_n^2 \bigr]^{1/2}$. We set
\begin{equation}\label{gkn'}
    G_{{k}'_n}^c (Q_n^a) = 1  \otimes \left[ -\Delta_{r_a} + Q_n^a \eta_{-1} (r_a) + {k'}_n^2 \right]^{-1}
\end{equation}
Acting on both sides of (\ref{schro:3})--(\ref{schro:4}) by $G_{{k}'_n}^c (Q_n^a) $ gives
\begin{equation}\label{schro:5}
P_{\varphi^a_i} (Z_n)  \psi_n  = - G_{{k}'_n}^c (Q_n^a) \eta_{-\lambda} (r_a)  P_{\varphi^a_i} (Z_n) \eta_{\lambda} (r_a)
    \left[ I_a^n - Q^a_n \eta_{-1} (r_a) \right] \psi_n ,
\end{equation}
where we have inserted $\eta_\lambda \eta_{-\lambda} = 1$.
Adding and subtracting $\phi_0$ from $\psi_n$ we rewrite (\ref{schro:5}) as inequality
\begin{gather}
\left| P_{\varphi^a_i} (Z_n) \psi_n \right| \\
\leq \left|  G_{{k}'_n}^c (Q_n^a)\eta_{-\lambda} (r_a)  P_{\varphi^a_i} (Z_n) \eta_{\lambda} (r_a) \left[ I^n_a - Q^a_n \eta_{-1} (r_a) \right] (\psi_n -\phi_0)\right| \\
+ \left|  G_{{k}'_n}^c (Q_n^a)  (Q_n^a)  \eta_{-\lambda} (r_a)P_{\varphi^a_i} (Z_n)\eta_{\lambda} (r_a) \left[ I^n_a - Q^a_n \eta_{-1} (r_a) \right] \phi_0\right|
\end{gather}
Applying Lemma~7 in \cite{cmp} we continue
\begin{gather}
\left| P_{\varphi^a_i} (Z_n) \psi_n \right| \\
\leq G_{k_n}^c (Q_0)  \eta_{-\lambda} (r_a) P_{|\varphi^a_i |} (Z_n)\Bigl|  \eta_{\lambda} (r_a)  \left[ I^n_a - Q^a_n \eta_{-1} (r_a) \right] (\psi_n -\phi_0)\Bigr| \\
+ G_{k_n}^c (Q_0)  \eta_{-\lambda} (r_a) P_{|\varphi^a_i |}  (Z_n) \Bigl|  \eta_{\lambda} (r_a)  \left[ I^n_a - Q^a_n \eta_{-1} (r_a) \right] \phi_0\Bigr|,
\end{gather}
where we define $P_{|\varphi^a_i |} := |\varphi^a_i | (\cdot, |\varphi^a_i |)\otimes 1$ and use $Q_n^a >Q_0$, ${k'}^2_n \geq k_n^2$.
Finally, applying Lemma~\ref{lem:thet} we write
\begin{equation}\label{gnhn}
    \left| P_{\varphi^a_i} (Z_n) \psi_n \right|  \leq g_n + P_{|\varphi^a_i |} (Z_n) e^{\beta |\xi_a|} |h_n |,
\end{equation}
where we define
\begin{gather}
g_n := G_{k_n}^c (Q_0)  \eta_{-\lambda} (r_a) P_{|\varphi^a_i |} (Z_n) e^{\beta |\xi_a|} \Bigl|  \Theta_a (x) (\psi_n -\phi_0)\Bigr| \label{gn}\\
|h_n| := G_{k_n}^c (Q_0)  \eta_{-\lambda} (r_a) \Bigl|  \Theta_a (x)\phi_0\Bigr| \label{hn}
\end{gather}
It remains to prove that both terms on the rhs of (\ref{gnhn}) do not spread. Then it would follow that $\left| P_{\varphi^a_i} (Z_n) \psi_n \right|$ does not spread and the statement would be proved. We have
\begin{gather}
\| g_n \| \leq \bigl\| G_{k_n}^c (Q_0)  \eta_{-\lambda} (r_a) \bigr\| \times \bigl\| P_{|\varphi^a_i |} (Z_n) e^{\beta |\xi_a|} \bigr\| \times \Bigl\|  \Theta_a (x) (\psi_n -\phi_0)\Bigr\|
\end{gather}
The first two operator norms are uniformly bounded by the Lemmas~\ref{lemma:2} and R7 (cf. Lemma~\ref{lem:4}). The last norm goes to zero because $\Theta_a (x) \in L^2(\mathbb{R}^{3N-3}) + L^\infty_\infty (\mathbb{R}^{3N-3})$ is relatively $H_0$ compact (see Lemma~3 in \cite{cmp}). Hence, $\|g_n\| \to 0$ and $g_n$ does not spread. By the same reasoning the sequence $|h_n|$ is uniformly norm--bounded. Following the arguments in the proof of Theorem~3 in \cite{cmp} after Eq.~(44) it is easy to show that $|h_n|$ satisfies all criteria of Theorem~1 in \cite{cmp} and, hence, does not spread. But then the sequence $P_{|\varphi^a_i |} (Z_n) e^{\beta|\xi_a|} |h_n|$ does not spread either by Lemma~\ref{lem:5} since $\| e^{\beta|\xi_a|} P_{|\varphi^a_i |} (Z_n) e^{\beta|\xi_a|} \| \leq A^2$. Thus $P^{(a)}_{thr} (Z_n) \psi_n $ does not spread and the claim is proved. \end{proof}

Now we can prove the main theorem.
\begin{proof}[of Theorem~\ref{th:main}]
Again, following the arguments in the proof of
Theorem~3 in \cite{cmp}, the theorem would be proved if we would show that every weakly converging subsequence of the sequence $\psi (Z_k)$ does not spread. So let $\{Z_n\}_{n=1}^\infty \subset \mathcal{Z}/Z_{cr}$, $Z_n \to Z_{cr}$ be a subsequence such that $\psi_n := \psi(Z_n )$ is weakly convergent: $\psi_n \wwto \phi_0$, where $\phi_0 \in D(H_0)$ by Lemma~3 in \cite{cmp}. From $\mathcal{P} \psi_n = \psi_n$ it is trivial to show that $\mathcal{P} \phi_0 = \phi_0$. Our aim is to show that $\psi_n$ does not spread.

The following identity is obvious
\begin{gather}
\psi_n = \sum_{a=1}^L P^{(a)}_{thr} (Z_n) J_a^2  (\psi_n -\phi_0) + \sum_{a=1}^L \bigl[ 1-P^{(a)}_{thr} (Z_n) \bigr] J_a^2
(\psi_n -\phi_0) \label{obvi} \\
+ \sum_{a=L+1}^{2^{N-1}-1}J_a^2
(\psi_n -\phi_0) + \phi_0 \label{obvi:1}
\end{gather}
The the last term in (\ref{obvi})--(\ref{obvi:1}) is a fixed $L^2$ function and the last two sums go to zero in norm by Lemma~\ref{lemma:3}. Hence, to show that $\psi_n$ does not spread it suffices to prove that each term in the first sum does not spread. Using $\sum_a J_a^2 =1$ we write
\begin{gather}
    \bigl| P^{(a)}_{thr} (Z_n) J_a^2  (\psi_n -\phi_0) \bigr| \leq \bigl| P^{(a)}_{thr}  (Z_n) (\psi_n - \phi_0)\bigr| + \sum_{b\neq a} \bigl| P^{(a)}_{thr}(Z_n) J_b^2 ( \psi_n - \phi_0 )\bigr| \leq \label{claim3:1}\\
         \bigl| P^{(a)}_{thr}  (Z_n) \psi_n \bigr| + \bigl| P^{(a)}_{thr}(Z_n) \phi_0 \bigr|  +
\sum_{b\neq a} \bigl| P^{(a)}_{thr}(Z_n) J_b^2 ( \psi_n - \phi_0 )\bigr|  \label{claim3:2}
\end{gather}
The first two term on the rhs of (\ref{claim3:2}) do not spread by Lemmas~\ref{lem:6},\ref{lem:5} respectively. If we would show that the sum on the rhs of (\ref{claim3:2}) goes to zero in norm then the statement of the theorem would follow. Indeed, for $b \neq a$
\begin{gather}
\bigl\| P^{(a)}_{thr}(Z_n) J_b^2 ( \psi_n - \phi_0 )\bigr\| \leq \bigl\| P^{(a)}_{thr}(Z_n) e^{\beta |\xi_a|} \bigr\| \times \bigl\|e^{-\beta |\xi_a|} J_b^2 ( \psi_n - \phi_0 )\bigr\|
\end{gather}
The operator norm is uniformly bounded by R7 and the second norm goes to zero because $e^{-\beta |\xi_a|} J_b^2 \in L^\infty_\infty (\mathbb{R}^{3N-3})$ for $a\neq b$ and is thus relatively $H_0$ compact (see Lemma~3 in \cite{cmp}). Thus
\begin{equation}\label{endlem}
\lim_{n \to \infty}    \bigl\| P^{(a)}_{thr}(Z_n) J_b^2 ( \psi_n - \phi_0 )\bigr\| = 0
\end{equation}
and the claim is proved.  \end{proof}

\section{Application to Atomic Ions}

Consider the Hamiltonian of an infinitely heavy atomic nucleus charge $Z$ containing $N$ electrons
\begin{equation}\label{atham}
    H_{N}(Z) = H_0 - \sum_{i=1}^{N} \frac Z{|x_i|} + \sum_{1 \leq i <j \leq N}\frac 1{|x_i - x_j|}
\end{equation}
The total number of particles is $N+1$ (the electrons are numbered from 1 to $N$ and the nucleus is the particle number $N+1$). By $\mathcal{P}_N$ we denote the projection operator on the proper symmetry subspace. We keep the integer number of electrons fixed and let the atomic charge $Z$ continuously vary. For $Z = N$ the Hamiltonian $H_{N}(Z)\mathcal{P}_N$ describes an atom and for $Z = N -1$ it describes a single negative atomic ion. We shall call the system described by the Hamiltonian (\ref{atham}) \emph{stable} if $H_{N}(Z)\mathcal{P}_N$ has a bound state below the bottom of the continuous spectrum. It is known \cite{zhislin} that for $Z = N$ the system is stable. When one diminishes $Z$ the nuclear charge eventually reaches the critical value $Z_{cr}$ such that for $Z> Z_{cr}$ the system is stable but for $Z\leq Z_{cr}$ already unstable. For a rigorous proof on existence of the critical charge see \cite{ruskai,sigalsays2}. It is a true mathematical challenge \cite{simonproblems} to prove that $Z_{cr} \geq N -2$.

For simplicity of the exposition we shall assume $Z_{cr} \in (N - 1 , N)$ (this is the case of some inert gases, for which no stable single negative ions exist \cite{hogreve}). In this case the whole proof can be done on a rigorous footing because the dissociation thresholds are known exactly. We can prove the following
\begin{theorem}\label{th:phys}
Suppose that $Z_{cr} \in (N - 1 , N)$. Then $H_{N}(Z_{cr}) \mathcal{P}_N$ has a bound state at the bottom of the continuous spectrum.
\end{theorem}
\begin{proof}
Let us first introduce the parameter sequence for $k = 1,2,\ldots$
\begin{equation}\label{zkdef}
Z_k = Z_{cr} + \frac{N-Z_{cr}}{2k}
\end{equation}
Obviously, $Z_k \to Z_{cr}$ and $Z_k \leq Z_1 < N$. Now our aim is to show that all conditions of Theorem~1 are fulfilled. It is helpful to denote by $E_N$ the ground state energy of $H_N \mathcal{P}_N$, or in other words to set $E_N := \inf \sigma (H_N \mathcal{P}_N)$. The requirements R1-2 are obviously fulfilled. R3 is fulfilled by the definition of the critical charge. Note that for $Z \in \mathcal{Z}$ the bottom of the continuous spectrum corresponds to the decay into one electron and the rest of the particles, that is $E_{thr} (Z) = E_{N-1} (Z)$. In the interval $Z \in (N-1, N)$ the function  $E_{N-1} (Z)$ is continuous and, hence, R4 is fulfilled. Let $a=1, \ldots, N$ number those partitions, where the electron number $a$ is removed from the rest of the particles. Then R5 is fulfilled, where one sets $L = N$. Eq.~(\ref{need1}) can be rewritten in the present terms as $E_{N-2} (Z ) \geq E_{N-1} (Z ) + 2 |\Delta \epsilon|$ for all $Z \in \mathcal{Z}$. This inequality would be satisfied if we set
\begin{equation}\label{need1text}
    2 |\Delta \epsilon| := \inf_{Z \in \mathcal{Z}} \bigl[ E_{N-2} (Z ) - E_{N-1} (Z ) \bigr] ,
\end{equation}
where obviously $|\Delta \epsilon| >0$. It is easy to see that with $|\Delta \epsilon|$ defined by (\ref{need1text}) the inequality (\ref{need2}) is automatically satisfied. Thus R6 is fulfilled. The only hard part is to check that R7 is satisfied.

Let us write (\ref{pthranew}) for $a = 1, \ldots, N$
\begin{equation}\label{pthranew2}
 P_{thr}^a (Z) = \sum_{i=1}^{n(Z)} E_i^a (Z) \varphi^a_i (\cdot, \varphi^a_i),
\end{equation}
where $E_i^a (Z) \in [E_{N-1}(Z) , E_{N-1}(Z) + |\Delta \epsilon|]$. By Lemma~\ref{lem:4} R7 would be satisfied if we would show that there exist $A, \beta >0$ independent of $Z$ such that $(\varphi^a_i (Z), e^{2\beta |\xi_a|} \varphi^a_i (Z)) \leq A $ for all $Z \in \mathcal{Z}$.

Recall that the coordinates $\xi^a_i$ for $a= 1,\ldots, N$ point from the nucleus to one of the $N-1$ electrons. Following \cite{ahlrichs} we use the inequality for the integer powers of $|\xi_a|$
\begin{equation}\label{polya}
|\xi_a|^n \leq (N-2)^{n-1} \sum_{i = 1}^{N-2} |\xi^a_i|^n
\end{equation}
to write
\begin{equation}\label{ahlr}
(\varphi^a_i, |\xi_a |^n\varphi^a_i) \leq (N-2)^n (\varphi^a_i, |\xi^a_1 |^n \varphi^a_i),
\end{equation}
where we have used the wave function being either symmetric or antisymmetric under permutation of electrons.
Representing the exponential function by series we get from (\ref{ahlr})
\begin{equation}\label{ahlr2}
(\varphi^a_i, e^{2\beta |\xi_a|} \varphi^a_i) \leq \sum_{n=0}^\infty \frac{(2\beta)^n}{n!} (N-2)^n (\varphi^a_i, |\xi^a_1 |^n \varphi^a_i)
\end{equation}
Now we use the inequality (2.20) from \cite{osten}
\begin{equation}\label{ahlr3}
    \frac{(\varphi^a_i, |\xi^a_1 |^{n+1} \varphi^a_i)}{(\varphi^a_i, |\xi^a_1 |^n \varphi^a_i)} \leq \frac 1{2\epsilon} \bigl\{ Z + [Z^2 + \frac{\epsilon}2 (n+2)^2]^{1/2} \bigr\} ,
\end{equation}
where $\epsilon := E_{N-2} (Z) - E_i^a (Z)$. Note that $\epsilon \geq |\Delta \epsilon|$.  For $Z \leq Z_1$ the inequality (\ref{ahlr3}) can be transformed into
\begin{equation}\label{ahlr4}
    (\varphi^a_i, |\xi^a_1 |^{n+1} \varphi^a_i) \leq (n+1)C(\varphi^a_i, |\xi^a_1 |^n \varphi^a_i),
\end{equation}
where we defined the constant $C$ through
\begin{equation}\label{ahlr5}
    C = \frac{Z_1}{2|\Delta \epsilon|} + \frac{1}{2|\Delta \epsilon|} \left( Z^2_1 + 2|\Delta \epsilon| \right)^{1/2}
\end{equation}
Using $\varphi^a_i$ being normalized from (\ref{ahlr4}) we get the upper bound
\begin{equation}\label{ahlr6}
 (\varphi^a_i, |\xi^a_1 |^n \varphi^a_i) \leq C^n n!
\end{equation}
that the lhs of (\ref{ahlr2}) would be bounded by a constant independent of $Z$ if we set
\begin{equation}\label{choose beta}
    \beta < \frac 1{2C(N-2)}
\end{equation}
Thus R7 is also satisfied. For $a = 1, \ldots, N$ we have $Q^a (Z) \geq Z_1 - (N-1) >0$, where $Z \in \mathcal{Z}$. Thus all requirements of Theorem~\ref{th:main} are fulfilled and the claim is proved.
\end{proof}

A few remarks are in order. For $N=2$ the statement of Theorem~\ref{th:phys} was conjectured in \cite{stil} (see also \cite{baker}) and proved by Thomas and Maria Hoffmann-Ostenhof together with Barry Simon in \cite{exist}. It is quite natural to extrapolate this result for arbitrary $N$, see \cite{hogreve}. The restriction $Z_{cr} \in (N - 1 , N)$ in the condition of Theorem~\ref{th:phys} is imposed in order to keep the proof completely rigorous (otherwise to apply Theorem~\ref{th:main} one would need additional assumptions concerning the nature of dissociation thresholds). In fact, the same result must hold for $Z_{cr} < N$.

\end{document}